\documentclass[conference,9pt,compsocconf]{IEEEtran}
\IEEEoverridecommandlockouts
\ifCLASSINFOpdf
\else
\fi
 \usepackage{setspace}
\usepackage{booktabs} % For formal tables
\usepackage[noadjust]{cite}
\usepackage{amsmath}
\usepackage{amssymb}
\usepackage{amsfonts}
\usepackage{float}
\usepackage{graphicx}
\usepackage{xcolor}
\usepackage{colortbl}
\usepackage{multirow}
\usepackage{mathtools}
\usepackage{algorithm}
\usepackage{algorithmicx}
\usepackage{algpseudocode}
\usepackage{tabto}
\usepackage{hhline}
\let\labelindent\relax
\usepackage{enumitem}
\usepackage{pifont}
\usepackage[none]{hyphenat}
\makeatletter
\renewcommand\footnoterule{%
  \kern-3\p@
  \hrule\@width 0.5\columnwidth
  \kern2.6\p@}
  \makeatother

\begin{document}
\sloppy
%\setlength\parskip{1pt}
%\linenumbers
\title{\LARGE{\textbf{Exploiting Vulnerabilities in Deep Neural Networks: \\Adversarial and Fault-Injection Attacks\vspace{-28pt}}}}
% -------------- Authors ---------------------------------%
\author{
 \IEEEauthorblockN{Faiq Khalid$^{1}$, 
       Muhammad Abdullah Hanif$^{1}$,
 		Muhammad Shafique$^{2}$}
  	\IEEEauthorblockA{$^1$\textit{Technische Universit\"at Wien (TU Wien), Vienna, Austria}\\
  	$^2$\textit{Division of Engineering, New York University Abu Dhabi (NYUAD), Abu Dhabi, United Arab Emirates}\\
  	Email: \{faiq.khalid,muhammad.hanif\}@tuwien.ac.at, muhammad.shafique@nyu.edu}
 }
\maketitle
%--------------------------------------------------------------------
\begin{abstract} 
From tiny pacemaker chips to aircraft collision avoidance systems, the state-of-the-art Cyber-Physical Systems (CPS) have increasingly started to rely on Deep Neural Networks (DNNs). However, as concluded in various studies, DNNs are highly susceptible to security threats, including adversarial attacks. In this paper, we first discuss different vulnerabilities that can be exploited for generating security attacks for neural network-based systems. We then provide an overview of existing adversarial and fault-injection-based attacks on DNNs. We also present a brief analysis to highlight different challenges in the practical implementation of the adversarial attacks. Finally, we also discuss various prospective ways to develop robust DNN-based systems that are resilient to adversarial and fault-injection attacks. 
\end{abstract}
\begin{IEEEkeywords}
Deep Neural Networks; Adversarial Attacks; Machine Learning Security; Fault-injection Attacks.\vspace{-4pt}
\end{IEEEkeywords}
%====================================================================
%====================================================================
\section{Introduction}\label{Introduction}
\begin{figure}[!t]
    \centering
    \includegraphics[width=1\linewidth]{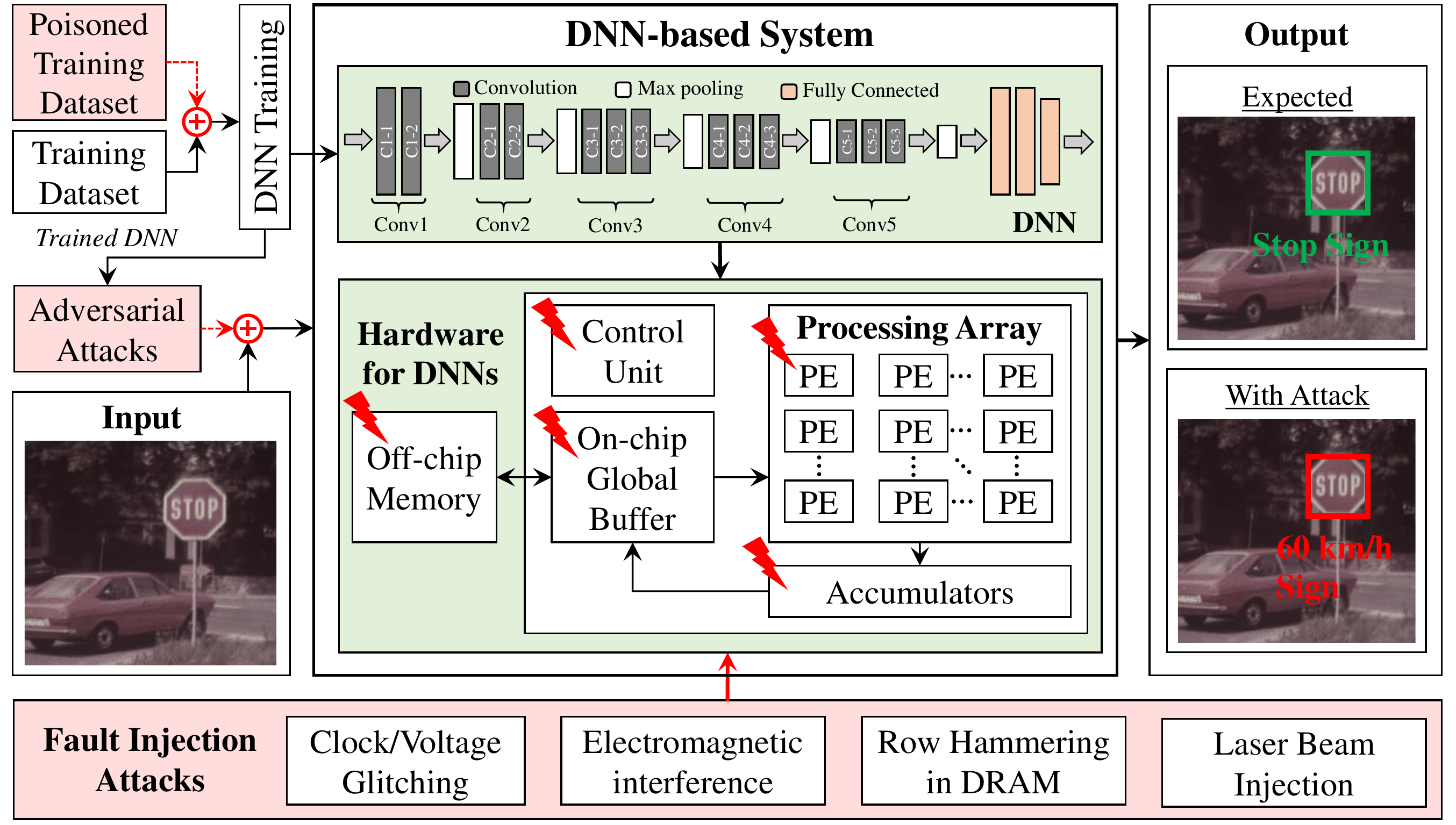}
    \caption{Security threats (i.e., adversarial and fault-injection attacks) to a DNN-based system. Fault-injection attacks can be performed by physically injecting faults in the off-chip and on-chip memories, and in the Processing Elements (PEs). The ``Stop'' sign image is take from the German Traffic Sign Detection dataset~\cite{Houben-IJCNN-2013}. }
    \label{fig:overview}
\end{figure}

Machine Learning (ML) algorithms have become popular in many applications, especially smart Cyber-Physical Systems (CPS), because of their ability to process and classify the enormous data~\cite{shafique2018overview,kriebel2018robustness,marchisio2019deep}, e.g., image recognition, object detection. The state-of-the-art ML systems are mostly based on Deep Neural Networks (DNNs), which consists of many layers of neurons connected into a mesh. The input follows this mesh from the input layer, through the hidden layers, to reach the output layer. The output node/neuron with the highest value indicates the decision of the DNN. However, due to several uncertainties in feature selection, ML algorithms inherently posses several security vulnerabilities, e.g., sensitivity for small input noise. Several attacks have been proposed to exploit these security vulnerabilities, for example, adversarial attacks~\cite{zhang2019building,hanif2018robust,shafique2020robust} (see Figure~\ref{fig:overview}). Depending upon the extent of the access to the training process, the DNN model, or the inference, the adversarial attacks can either exploit the input sensitivity of the trained DNN during the inference or manipulate the training process, DNN model or training dataset. 

Since the discovery of adversarial attacks, several studies have been performed but are mainly focused on the optimization and effectiveness of input data perturbation. In most of the practical cases, it is very challenging to manipulate the input of the DNN or corrupt the input data because of the limited access of DNN-based system or training process. However, the advancements in the communication network and computational elements of the CPS make us capable of putting the computing elements and sensors anywhere. This enables easy access to the computing elements and sensor, thereby opening a broader attack surface. Recently, this easy-to-get physical access is exploited to generate fault-injection attacks (see Figure~\ref{fig:overview}). In these attacks, attackers can externally inject the faults in data stored in the memory, the control path of a DNN-based system, or the computational blocks to manipulate the DNN output. These faults can be injected using well-known techniques, e.g., variations in voltage, Electromagnetic (EM) interference, and heavy-ion radiation.

\subsection{Novel Contributions}
To encompass the complete attack surface in the DNN-based system, in this paper, we study the security vulnerabilities with respect to adversarial attack and the emerging fault-injection attacks. \textbf{In summary, the contributions of this paper are:} 
\begin{enumerate}[leftmargin=*]
    \item We first study and discuss the different sets of assumptions and parameters of the threat model that can be used to show the effectiveness and practicability of an attack (\textbf{Section~\ref{sec:TM}}). 
    \item We discuss the different aspects, with respect to threat model, optimization algorithms, and computational cost, of the adversarial attacks (\textbf{Section~\ref{sec:AA}}) and fault-injection attacks (\textbf{Section~\ref{sec:FIA}}).  
    \item Most of the adversarial attacks do not consider the overall pipeline in the DNN-based system and ignore the pre-processing stages. Therefore, in this paper, we provide a brief analysis to highlight the challenges in the practical implementation of the adversarial attacks (\textbf{Section~\ref{sec:PIA}}). Based on this analysis, we conclude that the adversarial attack must be strong enough not to be nullified by the input pre-processing (which is low-pass filtering in our analysis). 
    \item Towards the end, we highlight different research directions on the road ahead towards developing a robust DNN-based system with stronger defenses against these attacks (\textbf{Section~\ref{sec:RD}}).   
\end{enumerate}

\section{Threat Model}\label{sec:TM}
\begin{figure}[!t]
	\centering
	\includegraphics[width=1\linewidth]{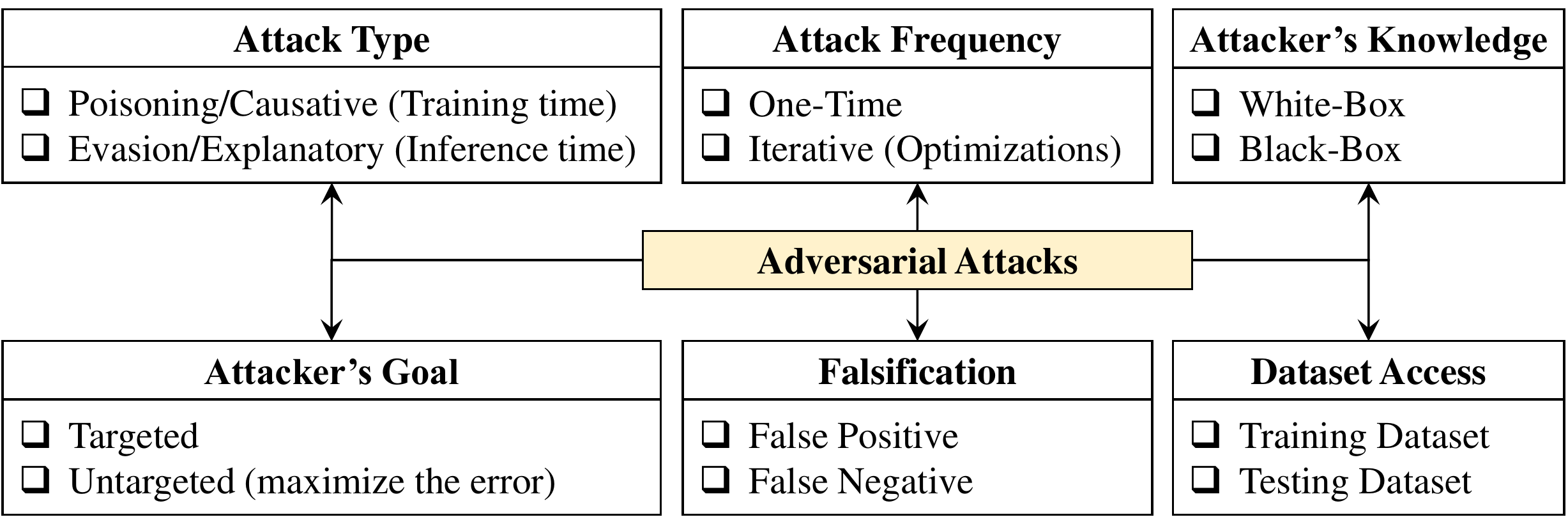}
	\caption{Threat model for adversarial attacks on DNNs. This model shows different assumptions and parameters that are required to generate an adversarial attack.}
	\label{fig:threat_model}
\end{figure}

\begin{figure}[h]
    	\centering
    	\includegraphics[width=1\linewidth]{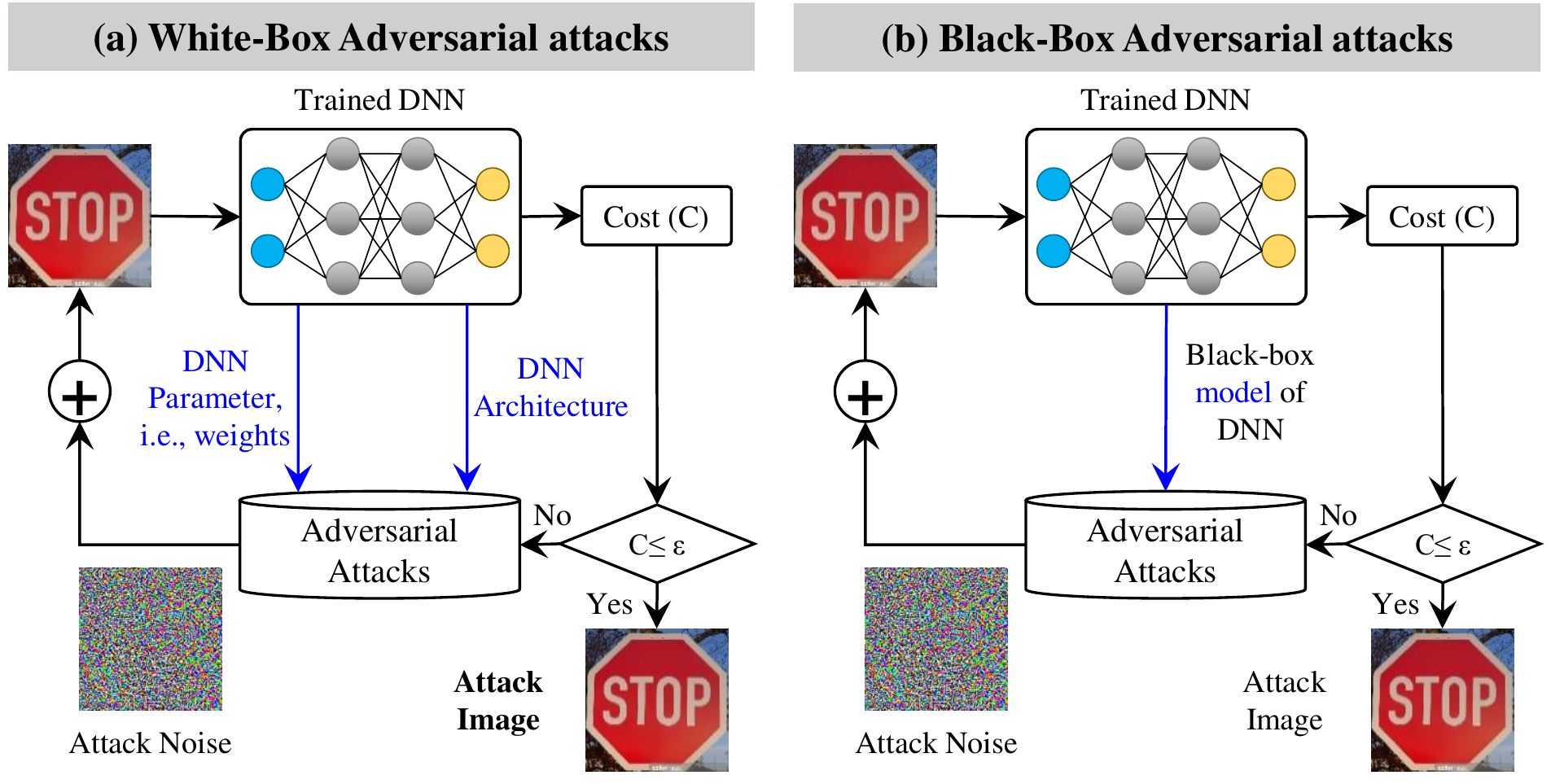}
    	\caption{An overview of the adversarial attack on machine learning. (a) In a white-box setting, the attacker has access to the network architecture and network parameters, i.e., weight, number of neurons, number of layers, activation functions, and convolution filters. (b) In a black-box setting, the attacker has access to the input and output of the network.}\vspace{-10pt}
    	\label{fig:advattack}
    \end{figure}

The effectiveness of an attack depend upon different assumptions and parameters. Different configurations of these assumptions and parameters generate several scenarios that are known as threat models. Therefore, in the context of DNN security, the threat model is based on the following set of parameters (see Figure~\ref{fig:threat_model}):  

\begin{itemize}[leftmargin=*]
    \item \textbf{Attacker's Knowledge:} Depending upon the information about the targeted ML system and attacker's access, the attack can either be black-box or white-box (see Figure~\ref{fig:advattack}). In \textit{white-box attacks}, the attacker has full access to the trained DNNs; thereby, allowing him to exploit DNN parameters to generate adversarial noise. In \textit{black-box attacks}, attacker has access only to the input and output of the DNN.
    \item \textbf{Attacker's Goal:} Depending upon the targeted payload, the attack can either be targeted or un-targeted. In the \textit{targeted attack}, the attacker modifies the DNN, input, or other parameters to achieve a particular misclassification. On the other hand, in the \textit{un-targeted attack}, the attacker's aim is only to maximize the prediction error.   
    \item \textbf{Attack Frequency:} To exploit a trade-off between resources, timing, and effectiveness, the attack can be wither one-shot or iterative. In \textit{one-shot}, the attack is optimized only once, but \textit{iterative} attack optimizes the payload of the attack over multiple iterations.  
    \item \textbf{Attack Falsification:} Depending upon the payload of the attack, these attacks can either be false positive or false negative attacks. In \textit{false-positive} attacks, a negative sample is misclassified as a positive sample. In \textit{false-negative} attacks, a positive sample is misclassified as a negative sample.  
    \item \textbf{Attack Type:} This parameter is related to the targeted phase of the ML design cycle. For example, depending upon the access, the attacker can target training, inference, or hardware of the DNN.  
    \item \textbf{Dataset Access:} The attack effectiveness and strength also depends upon the attacker's access to the different datasets. For example, the strength of the evasion attacks can significantly increase if the attacker has access to training and testing datasets. 
\end{itemize}

\section{Adversarial Attacks}\label{sec:AA}
The addition of imperceptible noise to the input can change the output of the DNN, and this phenomenon is known as the adversarial attack. It can also reduce the confidence classification that cause un-targeted misclassification, or it can also induce targeted misclassification. Several adversarial attacks have been proposed that manipulate the system to behave erroneously. Broadly, these attacks are categorized as causative and exploratory attacks. 

\begin{table*}[!t]
\footnotesize
\caption{A brief overview of the categorization of the state-of-the-art adversarial attacks on machine learning systems. (SVM: Support Vector Machines, SSI: Structural Similarity Index, and CC: Cross-correlation Coefficient}
\label{tab:advattack}
\resizebox{0.97\linewidth}{!}{
\begin{tabular}{lllllll}
\hline
\multicolumn{2}{|c|}{\textbf{Adversarial Attacks}} & \multicolumn{1}{l|}{\textbf{Type}} & \multicolumn{1}{l|}{\textbf{ Knowledge}} & \multicolumn{1}{l|}{\textbf{Frequency}} & \multicolumn{1}{l|}{\textbf{Goal}} & \multicolumn{1}{l|}{\textbf{Imperceptibility}} \\ \hline
\multicolumn{1}{|l|}{\multirow{19}{*}{\textbf{\begin{tabular}[c]{@{}l@{}}Gradient-\\ based\\ Attacks\end{tabular}}}} & \multicolumn{1}{l|}{Fast Gradient Sign Method (FGSM)\cite{goodfellow2014explaining}} & \multicolumn{1}{l|}{Evasion} & \multicolumn{1}{l|}{White-Box} & \multicolumn{1}{l|}{One-shot} & \multicolumn{1}{l|}{Targeted/Un-targeted} & \multicolumn{1}{l|}{$l_1$, $l_2$, $l_{inf}$ norms} \\ \cline{2-7} 
\multicolumn{1}{|l|}{} & \multicolumn{1}{l|}{Basic Iterative Method (BIM) or Iterative   FGSM\cite{kurakin2016adversarial}} & \multicolumn{1}{l|}{Evasion} & \multicolumn{1}{l|}{White-Box} & \multicolumn{1}{l|}{Iterative} & \multicolumn{1}{l|}{Targeted/Un-targeted} & \multicolumn{1}{l|}{$l_1$, $l_2$, $l_{inf}$ norms} \\ \cline{2-7} 
\multicolumn{1}{|l|}{} & \multicolumn{1}{l|}{Projected Gradient Descent (PGD)\cite{madry2017towards}} & \multicolumn{1}{l|}{Evasion} & \multicolumn{1}{l|}{White-Box} & \multicolumn{1}{l|}{Iterative} & \multicolumn{1}{l|}{Targeted/Un-targeted} & \multicolumn{1}{l|}{$\epsilon$ norm} \\ \cline{2-7} 
\multicolumn{1}{|l|}{} & \multicolumn{1}{l|}{Auto-PGD\cite{croce2020reliable}} & \multicolumn{1}{l|}{Evasion} & \multicolumn{1}{l|}{White-Box} & \multicolumn{1}{l|}{Iterative} & \multicolumn{1}{l|}{Targeted/Un-targeted} & \multicolumn{1}{l|}{$l_{inf}$ norm} \\ \cline{2-7} 
\multicolumn{1}{|l|}{} & \multicolumn{1}{l|}{Jacobian-based Saliency Map Attack (JSMA)\cite{papernot2016limitations}} & \multicolumn{1}{l|}{Evasion} & \multicolumn{1}{l|}{White-Box} & \multicolumn{1}{l|}{Iterative} & \multicolumn{1}{l|}{Targeted/Un-targeted} & \multicolumn{1}{l|}{$l_0$ norm} \\ \cline{2-7} 
\multicolumn{1}{|l|}{} & \multicolumn{1}{l|}{Iterative Frame Saliency\cite{inkawhich2018adversarial}} & \multicolumn{1}{l|}{Evasion} & \multicolumn{1}{l|}{White-Box} & \multicolumn{1}{l|}{Iterative/One-shot} & \multicolumn{1}{l|}{Un-targeted} & \multicolumn{1}{l|}{$l_0$ norm} \\ \cline{2-7} 
\multicolumn{1}{|l|}{} & \multicolumn{1}{l|}{Carlini \& Wagner $l_2$ attack\cite{carlini2017towards}} & \multicolumn{1}{l|}{Evasion} & \multicolumn{1}{l|}{White-Box} & \multicolumn{1}{l|}{Iterative} & \multicolumn{1}{l|}{Targeted/Un-targeted} & \multicolumn{1}{l|}{$l_2$ norm} \\ \cline{2-7} 
\multicolumn{1}{|l|}{} & \multicolumn{1}{l|}{Carlini \& Wagner $l_{inf}$ attack\cite{carlini2017towards}} & \multicolumn{1}{l|}{Evasion} & \multicolumn{1}{l|}{White-Box} & \multicolumn{1}{l|}{Iterative} & \multicolumn{1}{l|}{Targeted/Un-targeted} & \multicolumn{1}{l|}{$l_{inf}$ norm} \\ \cline{2-7} 
\multicolumn{1}{|l|}{} & \multicolumn{1}{l|}{DeepFool\cite{moosavi2016deepfool}} & \multicolumn{1}{l|}{Evasion} & \multicolumn{1}{l|}{White-Box} & \multicolumn{1}{l|}{Iterative} & \multicolumn{1}{l|}{Un-targeted} & \multicolumn{1}{l|}{$l_2$ norm} \\ \cline{2-7} 
\multicolumn{1}{|l|}{} & \multicolumn{1}{l|}{Universal Perturbations\cite{moosavi2017universal}} & \multicolumn{1}{l|}{Evasion} & \multicolumn{1}{l|}{White-Box} & \multicolumn{1}{l|}{Iterative} & \multicolumn{1}{l|}{Un-targeted} & \multicolumn{1}{l|}{$l_p$ norm} \\ \cline{2-7} 
\multicolumn{1}{|l|}{} & \multicolumn{1}{l|}{Newton Fool\cite{jang2017objective}} & \multicolumn{1}{l|}{Evasion} & \multicolumn{1}{l|}{White-Box} & \multicolumn{1}{l|}{Iterative} & \multicolumn{1}{l|}{Un-targeted} & \multicolumn{1}{l|}{Tuning parameter} \\ \cline{2-7} 
\multicolumn{1}{|l|}{} & \multicolumn{1}{l|}{Feature Adversaries\cite{sabour2015adversarial}} & \multicolumn{1}{l|}{Evasion} & \multicolumn{1}{l|}{White-Box} & \multicolumn{1}{l|}{Iterative} & \multicolumn{1}{l|}{Targeted} & \multicolumn{1}{l|}{$l_{inf}$ norm} \\ \cline{2-7} 
\multicolumn{1}{|l|}{} & \multicolumn{1}{l|}{Adversarial Patch\cite{brown2017adversarial}\cite{liu2019perceptual}} & \multicolumn{1}{l|}{Evasion} & \multicolumn{1}{l|}{White-Box} & \multicolumn{1}{l|}{Iterative} & \multicolumn{1}{l|}{Targeted/Un-targeted} & \multicolumn{1}{l|}{-} \\ \cline{2-7} 
\multicolumn{1}{|l|}{} & \multicolumn{1}{l|}{Elastic-Net (EAD)\cite{chen2018ead}} & \multicolumn{1}{l|}{Evasion} & \multicolumn{1}{l|}{White-Box} & \multicolumn{1}{l|}{Iterative} & \multicolumn{1}{l|}{Targeted/Un-targeted} & \multicolumn{1}{l|}{$l_1$, $l_2$, $l_{inf}$ norms} \\ \cline{2-7} 
\multicolumn{1}{|l|}{} & \multicolumn{1}{l|}{Dpatch\cite{liu2018dpatch}} & \multicolumn{1}{l|}{Evasion} & \multicolumn{1}{l|}{White-Box} & \multicolumn{1}{l|}{Iterative} & \multicolumn{1}{l|}{Targeted/Un-targeted} & \multicolumn{1}{l|}{-} \\ \cline{2-7} 
\multicolumn{1}{|l|}{} & \multicolumn{1}{l|}{High Confidence Low Uncertainty\cite{grosse2018limitations}} & \multicolumn{1}{l|}{Evasion} & \multicolumn{1}{l|}{White-Box} & \multicolumn{1}{l|}{Iterative} & \multicolumn{1}{l|}{Targeted} & \multicolumn{1}{l|}{$l_2$ norm} \\ \cline{2-7} 
\multicolumn{1}{|l|}{} & \multicolumn{1}{l|}{Waaserstein Attack\cite{wong2019wasserstein}} & \multicolumn{1}{l|}{Evasion} & \multicolumn{1}{l|}{White-Box} & \multicolumn{1}{l|}{Iterative} & \multicolumn{1}{l|}{Targeted/Un-targeted} & \multicolumn{1}{l|}{$l_p$ norm} \\ \cline{2-7} 
\multicolumn{1}{|l|}{} & \multicolumn{1}{l|}{Shadow Attack\cite{ghiasi2020breaking}} & \multicolumn{1}{l|}{Evasion} & \multicolumn{1}{l|}{White-Box} & \multicolumn{1}{l|}{Iterative} & \multicolumn{1}{l|}{Targeted/Un-targeted} & \multicolumn{1}{l|}{$l_2$, $l_{inf}$ norms} \\ \cline{2-7} 
\multicolumn{1}{|l|}{} & \multicolumn{1}{l|}{TrISec\cite{khalid2019trisec}} & \multicolumn{1}{l|}{Evasion} & \multicolumn{1}{l|}{White-Box} & \multicolumn{1}{l|}{Iterative} & \multicolumn{1}{l|}{Targeted/Un-targeted} & \multicolumn{1}{l|}{SSI, CR} \\ \hline
\multicolumn{1}{|l|}{\multirow{3}{*}{\textbf{\begin{tabular}[c]{@{}l@{}}Score-\\ based \\ Attacks\end{tabular}}}} & \multicolumn{1}{l|}{Zeroth Order Optimization (ZOO)\cite{chen2017zoo}} & \multicolumn{1}{l|}{Evasion} & \multicolumn{1}{l|}{Black-Box} & \multicolumn{1}{l|}{Iterative} & \multicolumn{1}{l|}{Targeted} & \multicolumn{1}{l|}{$l_2$ norm} \\ \cline{2-7} 
\multicolumn{1}{|l|}{} & \multicolumn{1}{l|}{Local Search\cite{narodytska2016simple}} & \multicolumn{1}{l|}{Evasion} & \multicolumn{1}{l|}{Black-Box} & \multicolumn{1}{l|}{Iterative} & \multicolumn{1}{l|}{Un-targeted} & \multicolumn{1}{l|}{-} \\ \cline{2-7} 
\multicolumn{1}{|l|}{} & \multicolumn{1}{l|}{Copy and Paste\cite{brunner2019copy}} & \multicolumn{1}{l|}{Evasion} & \multicolumn{1}{l|}{Black-Box} & \multicolumn{1}{l|}{Iterative} & \multicolumn{1}{l|}{Targeted/Un-targeted} & \multicolumn{1}{l|}{$l_2$ norm} \\ \hline
\multicolumn{1}{|l|}{\multirow{9}{*}{\textbf{\begin{tabular}[c]{@{}l@{}}Decision-\\ based \\ Attacks\end{tabular}}}} & \multicolumn{1}{l|}{HopskipJump\cite{chen2020hopskipjumpattack}} & \multicolumn{1}{l|}{Evasion} & \multicolumn{1}{l|}{Black-Box} & \multicolumn{1}{l|}{Iterative} & \multicolumn{1}{l|}{Targeted/Un-targeted} & \multicolumn{1}{l|}{$l_2$ norm} \\ \cline{2-7} 
\multicolumn{1}{|l|}{} & \multicolumn{1}{l|}{Query Efficient Attack \cite{cheng2018query}} & \multicolumn{1}{l|}{Evasion} & \multicolumn{1}{l|}{Black-Box} & \multicolumn{1}{l|}{Iterative} & \multicolumn{1}{l|}{Targeted/Un-targeted} & \multicolumn{1}{l|}{$l_2$ norm} \\ \cline{2-7} 
\multicolumn{1}{|l|}{} & \multicolumn{1}{l|}{Decision-based Attack \cite{brendel2017decision}} & \multicolumn{1}{l|}{Evasion} & \multicolumn{1}{l|}{Black-Box} & \multicolumn{1}{l|}{Iterative} & \multicolumn{1}{l|}{Targeted/Un-targeted} & \multicolumn{1}{l|}{$l_2$ norm} \\ \cline{2-7} 
\multicolumn{1}{|l|}{} & \multicolumn{1}{l|}{Query Efficient Boundary Attack (QEBA)\cite{li2020qeba}} & \multicolumn{1}{l|}{Evasion} & \multicolumn{1}{l|}{Black-Box} & \multicolumn{1}{l|}{Iterative} & \multicolumn{1}{l|}{Targeted/Un-targeted} & \multicolumn{1}{l|}{$l_2$ norm} \\ \cline{2-7} 
\multicolumn{1}{|l|}{} & \multicolumn{1}{l|}{Geometry-Inspired Decision-based (qFool)\cite{liu2019geometry}} & \multicolumn{1}{l|}{Evasion} & \multicolumn{1}{l|}{Black-Box} & \multicolumn{1}{l|}{Iterative} & \multicolumn{1}{l|}{Targeted/Un-targeted} & \multicolumn{1}{l|}{$l_2$ norm} \\ \cline{2-7} 
\multicolumn{1}{|l|}{} & \multicolumn{1}{l|}{Threshold Attack\cite{vargas2019robustness}} & \multicolumn{1}{l|}{Evasion} & \multicolumn{1}{l|}{Black-Box} & \multicolumn{1}{l|}{Iterative} & \multicolumn{1}{l|}{Targeted/Un-targeted} & \multicolumn{1}{l|}{$l_p$ norm} \\ \cline{2-7} 
\multicolumn{1}{|l|}{} & \multicolumn{1}{l|}{Square Attack\cite{andriushchenko2019square}} & \multicolumn{1}{l|}{Evasion} & \multicolumn{1}{l|}{Black-Box} & \multicolumn{1}{l|}{Iterative} & \multicolumn{1}{l|}{Targeted/Un-targeted} & \multicolumn{1}{l|}{$l_p$ norm} \\ \cline{2-7} 
\multicolumn{1}{|l|}{} & \multicolumn{1}{l|}{Pixel Attacks\cite{su2019one}\cite{vargas2019robustness}} & \multicolumn{1}{l|}{Evasion} & \multicolumn{1}{l|}{Black-Box} & \multicolumn{1}{l|}{Iterative} & \multicolumn{1}{l|}{Targeted/Un-targeted} & \multicolumn{1}{l|}{$l_p$ norm} \\ \cline{2-7} 
\multicolumn{1}{|l|}{} & \multicolumn{1}{l|}{FaDec\cite{khalid2019red}} & \multicolumn{1}{l|}{Evasion} & \multicolumn{1}{l|}{Black-Box} & \multicolumn{1}{l|}{Iterative} & \multicolumn{1}{l|}{Targeted/Un-targted} & \multicolumn{1}{l|}{$l_2$ norm, SSI, CR} \\ \hline
\multicolumn{1}{|l|}{\multirow{8}{*}{\textbf{\begin{tabular}[c]{@{}l@{}}Dataset\\ Poisoning\end{tabular}}}} & \multicolumn{1}{l|}{Poisoning Attack on SVM~\cite{xiao2015support}\cite{biggio2012poisoning}} & \multicolumn{1}{l|}{Poisoning} & \multicolumn{1}{l|}{White-Box} & \multicolumn{1}{l|}{Iterative} & \multicolumn{1}{l|}{Targeted} & \multicolumn{1}{l|}{-} \\ \cline{2-7} 
\multicolumn{1}{|l|}{} & \multicolumn{1}{l|}{Targeted Clean-Label Poisoning~\cite{zhu2019transferable}} & \multicolumn{1}{l|}{Poisoning} & \multicolumn{1}{l|}{White-Box} & \multicolumn{1}{l|}{One-shot} & \multicolumn{1}{l|}{Targeted} & \multicolumn{1}{l|}{-} \\ \cline{2-7} 
\multicolumn{1}{|l|}{} & \multicolumn{1}{l|}{Watermarking~\cite{shafahi2018poison}} & \multicolumn{1}{l|}{Poisoning} & \multicolumn{1}{l|}{White-Box} & \multicolumn{1}{l|}{One-shot} & \multicolumn{1}{l|}{Targeted} & \multicolumn{1}{l|}{-} \\ \cline{2-7} 
\multicolumn{1}{|l|}{} & \multicolumn{1}{l|}{Efficient Dataset Poisoning~\cite{shafahi2018poison}} & \multicolumn{1}{l|}{Poisoning} & \multicolumn{1}{l|}{White-Box} & \multicolumn{1}{l|}{Iterative} & \multicolumn{1}{l|}{Targeted/Un-targeted} & \multicolumn{1}{l|}{-} \\ \cline{2-7} 
\multicolumn{1}{|l|}{} & \multicolumn{1}{l|}{BadNets~\cite{gu2019badnets}} & \multicolumn{1}{l|}{Poisoning} & \multicolumn{1}{l|}{White-Box} & \multicolumn{1}{l|}{Iterative} & \multicolumn{1}{l|}{Targeted} & \multicolumn{1}{l|}{-} \\ \cline{2-7} 
\multicolumn{1}{|l|}{} & \multicolumn{1}{l|}{Targeted Backdoor~\cite{chen2017targeted}} & \multicolumn{1}{l|}{Poisoning} & \multicolumn{1}{l|}{White-Box} & \multicolumn{1}{l|}{Iterative} & \multicolumn{1}{l|}{Targeted} & \multicolumn{1}{l|}{-} \\ \cline{2-7} 
\multicolumn{1}{|l|}{} & \multicolumn{1}{l|}{Dynamic Backdoor Attacks~\cite{salem2020dynamic}} & \multicolumn{1}{l|}{Poisoning} & \multicolumn{1}{l|}{White-Box} & \multicolumn{1}{l|}{Iterative} & \multicolumn{1}{l|}{Targeted} & \multicolumn{1}{l|}{-} \\ \cline{2-7} 
\multicolumn{1}{|l|}{} & \multicolumn{1}{l|}{Feature Collision Attack~\cite{shafahi2018poison}} & \multicolumn{1}{l|}{Poisoning} & \multicolumn{1}{l|}{White-Box} & \multicolumn{1}{l|}{Iterative} & \multicolumn{1}{l|}{Targeted/Un-targeted} & \multicolumn{1}{l|}{-} \\ \hline
\multicolumn{1}{|l|}{\multirow{2}{*}{\textbf{\begin{tabular}[c]{@{}l@{}}Model \\ Poisoning\end{tabular}}}} & \multicolumn{1}{l|}{Weight poisoning~\cite{kurita2020weight}} & \multicolumn{1}{l|}{Poisoning} & \multicolumn{1}{l|}{White-Box} & \multicolumn{1}{l|}{One-shot} & \multicolumn{1}{l|}{Targeted/Un-targeted} & \multicolumn{1}{l|}{-} \\ \cline{2-7} 
\multicolumn{1}{|l|}{} & \multicolumn{1}{l|}{Local Model Poisoning~\cite{fang2019local}} & \multicolumn{1}{l|}{Poisoning} & \multicolumn{1}{l|}{White-Box} & \multicolumn{1}{l|}{One-shot} & \multicolumn{1}{l|}{Targeted/Un-targeted} & \multicolumn{1}{l|}{-} \\ \hline
\end{tabular}}
\vspace{-10pt}
\end{table*}

\subsection{Evasion (Exploratory) Attacks}
In these attacks, an attacker introduces an imperceptible noise at the input of the trained DNN during the inference. This imperceptible noise (known as adversarial noise) can either perform targeted misclassification or maximize the prediction error. Since these attacks explore vulnerabilities of the trained DNN during inference, therefore, these attacks are also known as exploratory attacks. On the basis of how the attacker implements the adversarial attack, the attacks can be divided into three categories: \textit{gradient-based}, \textit{score-based} and \textit{decision-based} attacks, as shown in Table~\ref{tab:advattack}.  

\subsubsection{Gradient-based Attacks}
The gradient-based attacks make use of the network gradients to craft the attack noise. There are several gradient-based attacks that utilize different optimization algorithms and imperceptibility parameters to improve the effectiveness of these attacks (see the summary of gradient-based attacks in Table~\ref{tab:advattack}). However, most of them are based basic gradient-based attacks, i.e., Fast Gradient Sign Method (FGSM)~\cite{goodfellow2014explaining}, iterative-FGSM (iFGSM)~\cite{kurakin2016adversarial}, Jacobian Saliency Map Attack (JSMA)~\cite{papernot2016limitations},  Carlini and Wagner (C\&W)~\cite{carlini2017towards} attack and training dataset unaware attack (TrISec)~\cite{khalid2019trisec}. In this section, we only discuss these basic gradient-based attacks. 
\begin{itemize}[leftmargin=*]
    \item \textit{Fast Gradient Sign Method (FGSM)}~\cite{goodfellow2014explaining} is a gradient-based attack that utilizes use of the cost function $J(p, x, y_{True})$ for the network with parameters $p$, input $x$ and the correct output class $y_{True}$ to determine the direction in which the adversarial noise will have the greatest impact. Afterwards, a small noise $\epsilon$ is added in a single iteration to the input, in the direction of the obtained gradient. 
    
    \item The \textit{Iterative Fast Sign Method (iFGSM)}~\cite{kurakin2016adversarial} is a variant of FGSM that is preferred for targeted misclassification. Here, the cost function chosen corresponds to a specific target (misclassification) class. Instead of perturbing the input in one step, the input is perturbed by $\alpha$ in each step.
    
    \item In \textit{Jacobian Saliency Map Attack (JSMA)}~\cite{papernot2016limitations}, initially the forward derivative, i.e., the Jacobian, of the network $F$ with respect to all input nodes is determined. Accordingly, saliency map is constructed, which highlights the inputs that are the most vulnerable to noise. This map can be used to perturb the minimum number of inputs to implement a successful attack. A vulnerable input node is then perturbed towards the target class. If the perturbation is found to be insufficient, then the whole process of finding Jacobian and constructing the saliency map is repeated until a successful attack is implemented or the perturbation added to the input exceeds a given threshold $\gamma$. 
    
    \item The \textit{Carlini and Wagner (C\&W) attack}~\cite{carlini2017towards} is a white-box approach to adversarial attack. The main idea is to consider finding the appropriate perturbation for the adversarial attack as a dual optimization problem. The first objective is to minimize the noise with respect to $l_p$ norm, added to an image with the objective of obtaining a specific target label at output. The second objective is to minimize $f(x+\epsilon)$ to a non-positive value, where the output label of the input changes to the targeted misclassification label. 
    
    \item \textit{TrISec}~\cite{khalid2019trisec} proposes a training data ``unaware'' adversarial attack. The attack is again modeled as an optimization problem. Backpropagation is used to obtain the noise that causes targeted misclassification while minimizing the cost function associated with the network. The probability $P(F(x+\epsilon) = y_{Target})$ of the target class is simultaneously maximized to ensure a minimal and scattered perturbation based on two parameters, i.e., Cross-correlation Coefficient (CC) and Structural Similarity Index (SSI). CC ensures that the noise added to the input is imperceptible (to humans), while SSI ensures that noise is spread across the whole input instead of being focused in a small region, hence, improving the imperceptibility withe respect to subjective analysis.
\end{itemize}
Note, these attacks require the attacker to have access to the network's internal parameters to calculate the gradients and hence are generally carried out in a white-box environment. Note, all the white-box attacks can be implemented in the black-box environment when they are combined with model stealing attacks. These attacks are known as substitute model attacks. However, it is not always possible to have the white-box access to the trained DNN.  

\subsubsection{Score-based Evasion Attacks}
In these attacks, the attacker has access to output scores/probabilities~\cite{chen2017zoo}\cite{narodytska2016simple}. The generation of attack is formulated as an optimization problem where the change in output scores due to the input manipulation is used to predict the direction and strength of the next input manipulation. 

\begin{figure*}[!t]
	\centering
	\includegraphics[width=1\linewidth]{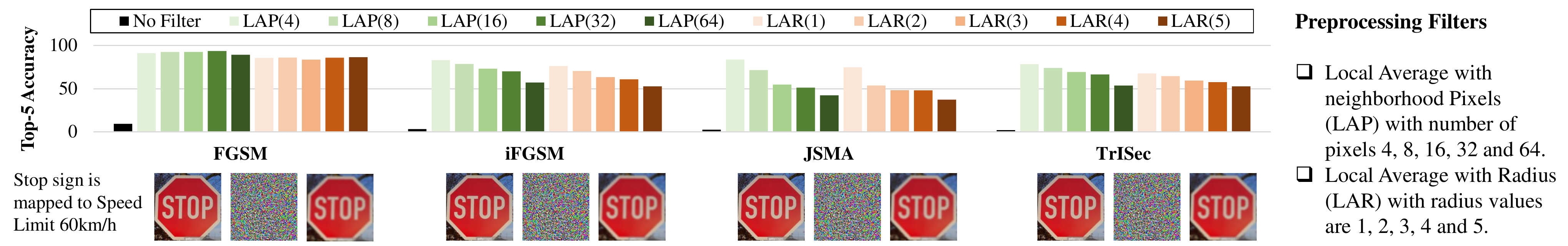}
	\caption{Effects of the preprocessing filters on some white-box adversarial attacks. It can be observed that introducing a simple low-pass filter can significantly impact the robustness of the adversarial attack. For example, in all attacks, the top-5 accuracy increases significantly, i.e., 10\% to 89\% with the addition of input preprocessing. However, as the smoothing factor of the filters surpasses a certain threshold, the top-5 accuracy started to decrease. The reason behind this behavior is that after this threshold, the filters started to affect the important features.\vspace{-15pt}}
	\label{fig:results}
\end{figure*}

\subsubsection{Decision-based Evasion Attacks}
These attacks make input alterations in a reverse direction - the procedure starts with a larger input noise that causes output misclassification. The manipulations are then iteratively reduced until they are imperceptible, while still triggering the adversarial attack~\cite{chen2020hopskipjumpattack,cheng2018query,brendel2017decision,li2020qeba,liu2019geometry,vargas2019robustness,andriushchenko2019square,su2019one}. Since these attack aims to estimate the adversarial example at the classification boundary, therefore, these attacks are also known as boundary attacks. In these attacks, the attack starts from a seed input from the target class (in case of a targeted attack) or any other incorrect output class (for random misclassification). The algorithm progresses iteratively towards the decision boundary of the true output class for input under attack. The objective is not only to reach the decision boundary, but also to explore the different parts of the boundary to ensure that a minimum amount of noise is being added to the input. Hence, no knowledge of the DNN's gradients, parameters, or output scores is required for a successful attack. However, the cost of these attacks in terms of the number of queries is very large. 

To reduce the number of queries, a \textit{Resource Efficient Decision-based attack (FaDec-attack)}~\cite{khalid2019red} finds targeted and un-targeted adversarial perturbations at a reduced computational cost. Like the boundary attack, a seed input from an incorrect class is chosen. The seed is iteratively modified to minimize its distance to the classification boundary of the original input. To reduce the computational cost of these attacks, adaptive step sizes are used to reach the smallest perturbation in the least number of iterations.

\subsection{Poisoning (Causative) Attacks}
In these attacks, the attacker manipulates the training algorithm, un-trained model, training dataset to influence, or corrupt the ML model itself. Based on the targeted components of the training process, these attacks can be categorized as dataset poisoning and model poisoning attacks (see Table~\ref{tab:advattack}).   
\begin{itemize}[leftmargin=*]
    \item \textit{Dataset Poisoning:} In these attacks, attacker manipulate the training dataset by adding patches (tailored noise) or randomly distributed noise~\cite{xiao2015support,biggio2012poisoning,zhu2019transferable,shafahi2018poison,jagielski2018manipulating,gu2019badnets,chen2017targeted}. These poisoned images influence different parameters of the DNN model such that it performs either target misclassification or maximizes the classification error. Since these attacks introduce the backdoors in the trained network that can be exploited during inference, therefore, these attacks are also known as backdoor attacks. 
    \item \textit{Model Poisoning:} Another type of causative attack is to slightly modify the DNN architecture such that for a particular trigger, it performs either target misclassification or maximizes the classification error~\cite{kurita2020weight}\cite{fang2019local}.
\end{itemize}

\section{Practical Implication of Evasion Attacks}\label{sec:PIA}
In adversarial attacks, the attack noise must be large enough to be captured by the acquisition device (for instance, camera) but imperceptible to the subjective analysis (by a human observer). 
A complete pipeline of the DNN-based classifier consists of an input sensor (for instance, a camera), a preprocessing module (e.g., filters), and a classifier. The robustness of the adversarial attacks depends upon the attacker's access to the different parts of the pipeline stages. 

\begin{figure}[h]
	\centering
	\includegraphics[width=1\linewidth]{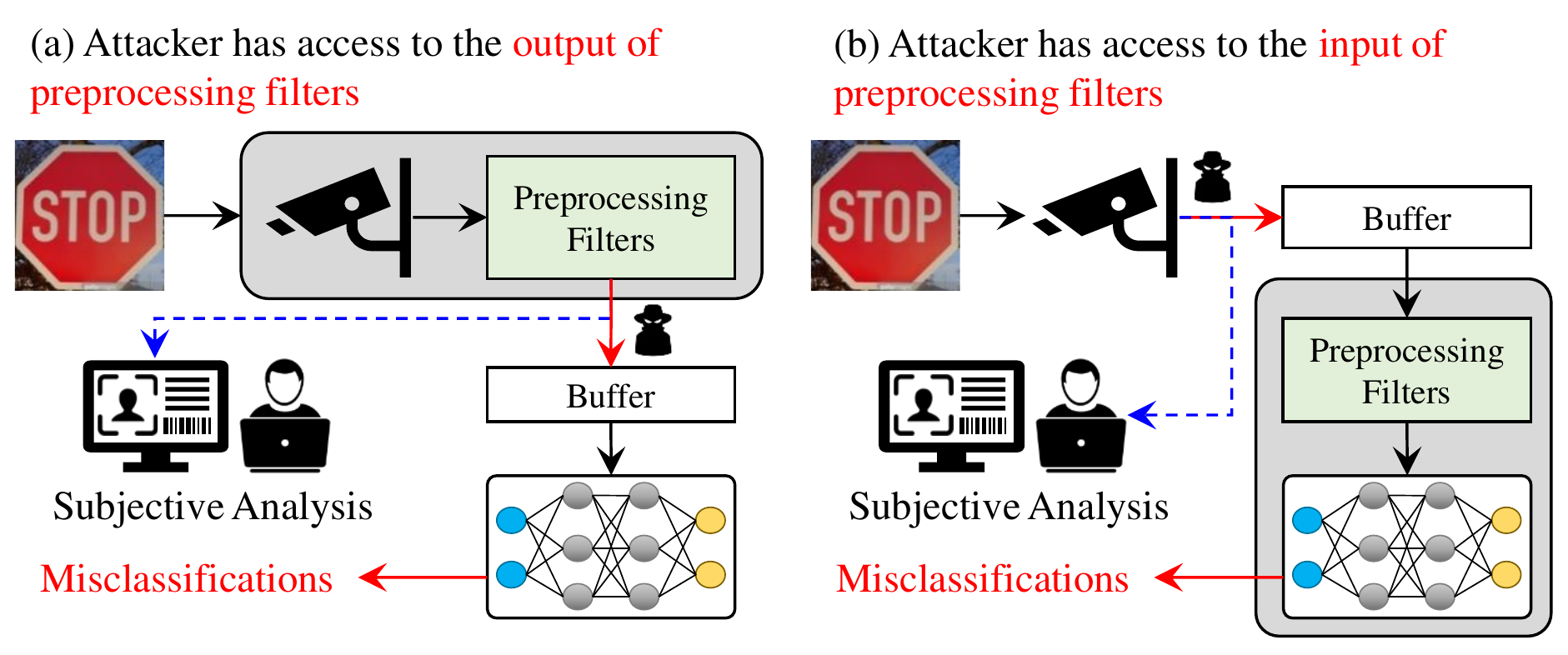}
	\caption{Practical implications of the adversarial attacks with respect to the attacker access. (a) Attack Model I: the attacker has access to the output of preprocessing filters. (b) Attack Model II: the attacker has access to the input of preprocessing filters. Note, for a successful attack, the adversarial noise should be robust to environmental changes.}
	\label{fig:PI}
\end{figure}

Traditionally, the adversarial attacks assumed that the attacker has access to the pre-processed input, as shown in Figure~\ref{fig:PI}(a). However, in real-world scenarios, it is very difficult to get access to the output of the preprocessing. The more realistic attack model considers the attacker to have access to the input of the preprocessing module (see Figure~\ref{fig:PI}(b)), for example, the camera is compromised, and it is generating and adding the adversarial noise. To analyze the impact of attack model II, we perform an analysis with low pass filters as a preprocessing module for basic white-box adversarial attacks, i.e., FGSM, JSMA, iFGSM, and TrISec. In this analysis, we choose the two commonly used noise filters: \textit{Local Average with neighborhood Pixels} (LAP) and \textit{Local Average with Radius} (LAR). Figure~\ref{fig:results} shows the impact of preprocessing filters on the adversarial attack and the key observations from the analysis are as follows:   
\begin{enumerate}[leftmargin=*]
    \item In the case of attack model II, the filters significantly reduces the effectiveness of all the implemented adversarial attacks.
    \item The increase in the number of neighboring pixels in the LAP filter worsens the performance of the DNN because it affects the key features of the input. Similarly, an increase in the LAR filter radius debilitates the performance of the DNN. 
\end{enumerate}

Based on the above-mentioned observations, in the context of adversarial attacks, we identify the following research directions:
\begin{enumerate}[leftmargin=*]
    \item To increase the robustness of adversarial attacks, it is imperative to incorporate the effects of preprocessing modules. For example, recently, researchers have presented that by incorporating the effects of preprocessing filters in optimization algorithms of existing adversarial attacks, the robustness of these attacks can be increased~\cite{khalid2019fademl}. Although this analysis considers only white-box attacks, it can effectively be extended to black-box attacks. 
    \item On the other hand, under a particular attack setting, preprocessing modules can also be used to nullify the adversarial attacks. For example, quantization~\cite{khalid2019qusecnets}, Sobel filters~\cite{ali2019sscnets} and transformations~\cite{raff2019barrage} are used to defend against adversarial attacks. However, the scope of these defenses is very limited, and most of them are breakable using black-box attacks. Therefore, it is a dire need to explore the practical implications of the adversarial attack to develop more powerful defenses.  
\end{enumerate}

\section{Fault-Injection Attacks}\label{sec:FIA}

Similar to adversarial attacks where the input to a DNN is modified to achieve misclassification, network parameters or computations can also be modified to achieve the same goal. Fault-injection attacks on DNNs refer to the attacks where an attacker tries to manipulate the output of a DNN by injecting faults in its parameters or in the data or control path of the hardware. There are several techniques that can be used for injecting faults, e.g., variations in voltage/clock signal, EM interference and heavy-ion radiation. 

% Threat Model: White-box attack. The attacker has knowledge of targeted DNN structure, its parameters and other low-level implementation details such as the location of the parameters in memory. 

\begin{figure}[h]
	\centering
	\includegraphics[width=1\linewidth]{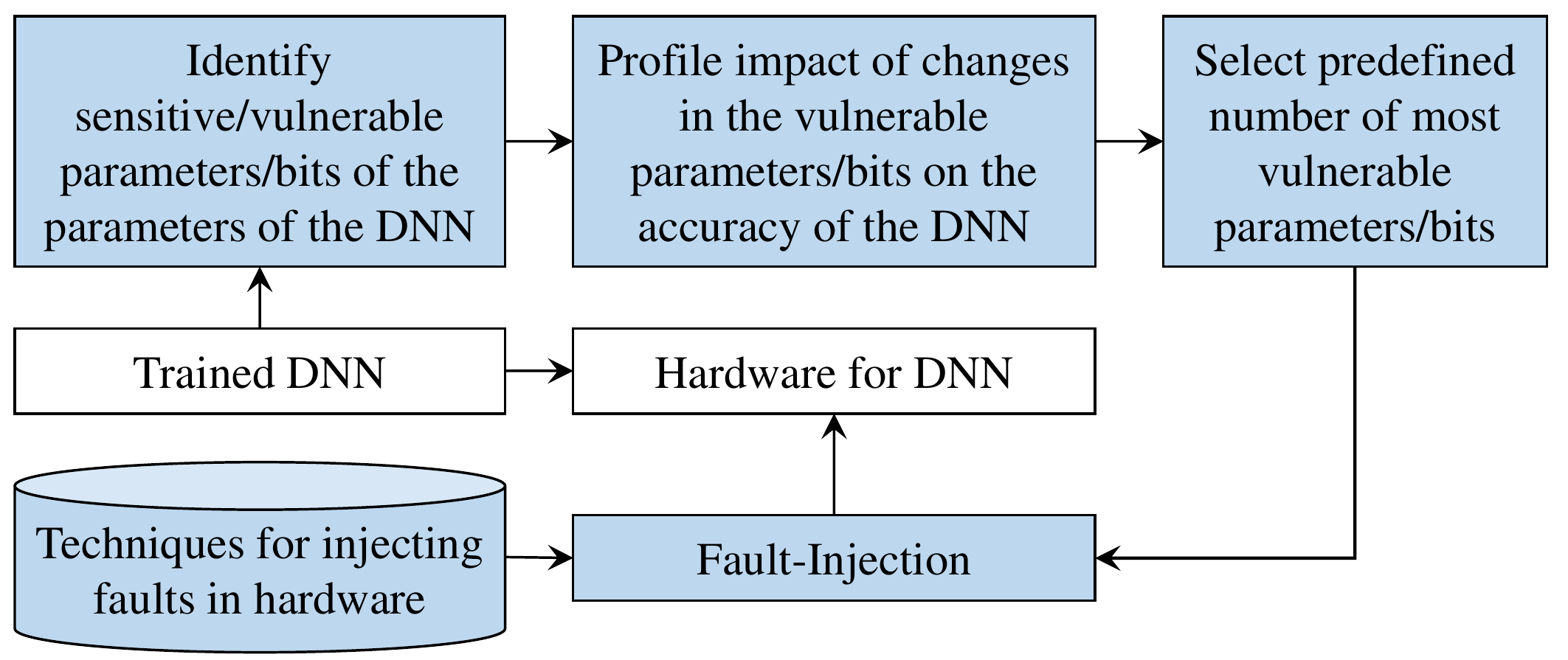}
	\caption{An overview of the design methodology for fault-injection attacks on ML-based systems.}
	\label{fig:flow_FIA}
\end{figure}

Several studies have been conducted towards modifying the network parameters to attack DNNs. Liu et al. in~\cite{liu2017fault} proposed two fault injection attacks, i.e., Single Bias Attack (SBA) and Gradient Descent Attack (GDA). Since the output of the DNNs is highly dependent on the biases in the output layer, SBA is realized by increasing only a single bias value corresponding to the neuron designated for the adversarial class. SBA is designed for the cases where the stealthiness of the attack is not necessary. For cases where stealthiness is important, GDA has been proposed that uses gradient descent to find the set of parameters to be modified and applies modifications to only some selected ones to minimize the impact of the injected faults on input patterns other than the specified one. Along the same direction, Zhao et al. in~\cite{zhao2019fault} proposed fault sneaking attack where they apply Alternating Direction Method of Multipliers (ADMM)~\cite{liu2019linearized} to optimize the attack while ensuring that the classification of images other than the ones specified is unaffected and the modification in the parameters is minimum. A generic flow of the fault-injection attacks on DNNs is shown in Figure~\ref{fig:flow_FIA}. 

\begin{figure}[h]
	\centering
	\includegraphics[width=1\linewidth]{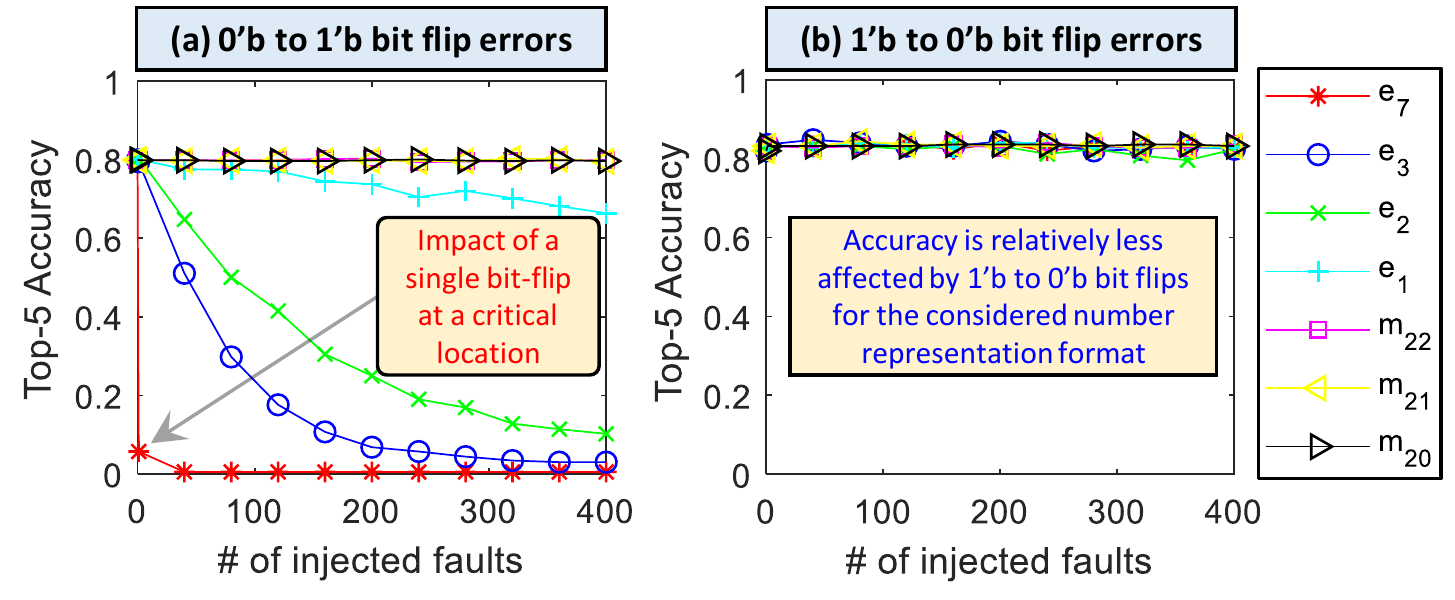}
	\caption{Impact of bit-flips in the weights of the VGG-f network on its classification accuracy on ImageNet dataset~\cite{hanif2018robust}. Figure~\ref{fig:bit_representation} shows the number representation format used for weights and activations for the analysis.}\vspace{-10pt}
	\label{fig:bit_flips}
\end{figure}

\begin{figure}[!t]
	\centering
	\includegraphics[width=1\linewidth]{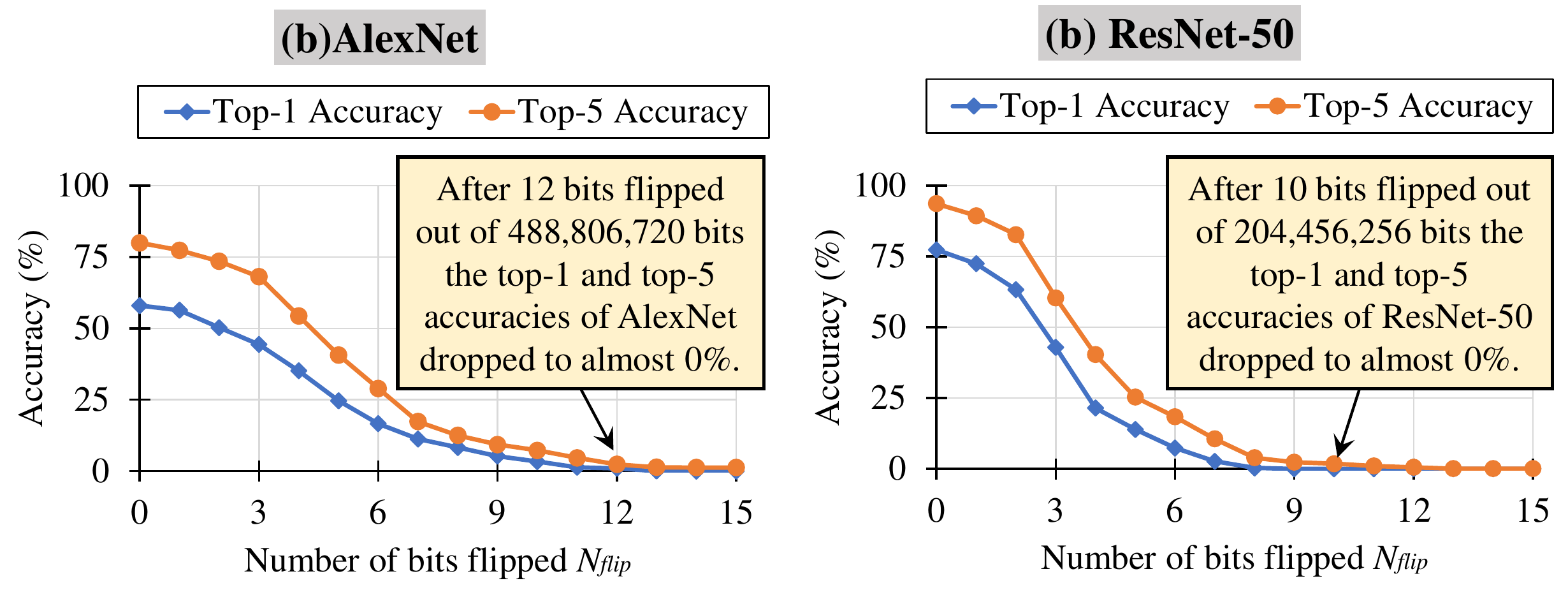}
	\caption{Accuracy vs. the number of bit-flips ($N_{flip}$) under BFA, for the AlexNet and the ResNet-50 on the ImageNet dataset (data source:~\cite{rakin2019bit}).}
	\label{fig:BF_results}
\end{figure}

To increase the stealthiness of the attack, Rakin et al. in~\cite{rakin2019bit} proposed a methodology, Bit-Flip Attack (BFA), for attacking DNNs by flipping a small number of bits in the weights of the DNN. The bit-flips can be performed through Row-Hammer attack~\cite{kim2014flipping}\cite{razavi2016flip} or laser injection~\cite{selmke2015precise}\cite{agoyan2010flip} when the weights are in the DRAM or the SRAM of the system, respectively. BFA focuses on identifying the most vulnerable bits in a given DNN that can maximize the accuracy degradation while requiring a very small number of bit-flips in the binary representation of the parameters of the DNN. It employs gradient ranking and progressive search to locate the most vulnerable bits. It is designed for quantized neural networks, i.e., where the weight magnitude is constrained based on the fixed-point representation. For floating-point representation, even a single bit-flip at the most significant location of the exponent of one of the weights of the DNN can result in the network generating completely random output~\cite{hanif2018robust} (see Figure~\ref{fig:bit_flips}). The results showed that BFA causes the ResNet-18 network to generate completely random output with only 13 bit-flips on the ImageNet dataset. Figure~\ref{fig:BF_results} presents the results for the AlexNet and the ResNet-50 networks under BFA.

Studies have also been conducted for injecting faults in the computations during the execution of the DNNs. Towards this, Breier et al. in~\cite{breier2018practical} performed an analysis of using a laser to inject faults during the execution of activation functions in a DNN to achieve misclassification. They focused on the instruction skip/change attack model, as it is one of the most basic (and repeatable) attacks for microcontrollers~\cite{breier2015laser}, to target four different activation functions, i.e., ReLU, sigmoid, Tanh and softmax. Batina et al. in~\cite{batina2018csi} showed that it is possible to reverse engineer a neural network using a side-channel analysis, e.g., by measuring the power consumption of the device during the execution of a DNN. Therefore, the attack proposed by Breier et al. can be employed even when the DNN is unknown.

\begin{figure}[!t]
	\centering
	\includegraphics[width=0.7\linewidth]{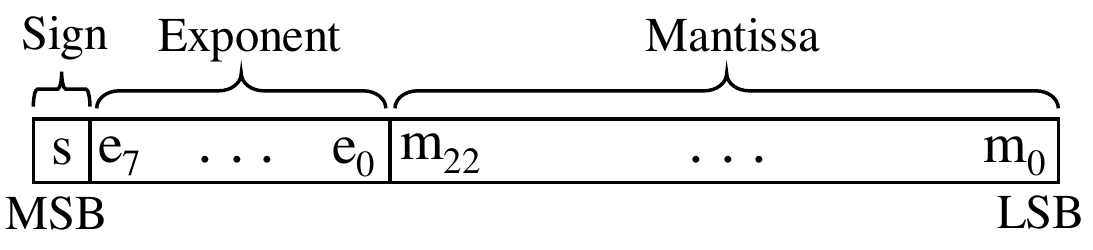}
	\caption{Single-precision floating-point representation}
	\label{fig:bit_representation}
\end{figure}

\section{Research Directions}\label{sec:RD}
Although DNNs are rapidly evolving and becoming an integral part of the decision-making process in CPS. However, the security vulnerabilities of DNNs, e.g., adversarial and fault-injection attacks, raises several concerns regarding their use in CPS. Therefore, stronger defenses against these attacks are required. Towards this, we identify some of the critical research directions: 
\begin{enumerate}[leftmargin=*]
    % \item There are several defenses exist to counter the adversarial attacks~\cite{khalid2019qusecnets,ali2019sscnets,khalid2019fademl}. However, all the available countermeasures for adversarial attacks are effective only for a particular type of adversarial attack. This calls for a deeper understanding of ``when" a system fails to the adversarial attack, to enable the design of an optimal defense against the attack.
    
    \item Several defenses have been proposed to counter the adversarial attacks~\cite{khalid2019qusecnets,ali2019sscnets,khalid2019fademl}. However, all the available countermeasures are effective only against a particular type of adversarial attacks. This calls for a deeper understanding of the existing attacks, to enable the design of an optimal defense. 
    
    % \item The security measures generally proposed for adversarial attacks are either applicable to a certain subclass of adversarial attacks or provide a defense against only the known attacks. Therefore, formal verification of DNNs is a promising approach to ensure adversarial robustness of DNN~\cite{shafique2020robust,naseer2020fannet} and thus can be used to develop verifiable security measures. The sound mathematical reasoning of formal verification techniques can provide complete and reliable security guarantees against adversarial attacks in DNNs.
    
    \item The security measures proposed for adversarial attacks are mainly effective against only a subclass of these attacks. Therefore, formal verification of DNNs is emerging as a promising approach to ensure adversarial robustness of DNNs~\cite{shafique2020robust}\cite{naseer2020fannet}, which can also help in developing verifiable security measures. The sound mathematical reasoning of formal verification techniques can provide complete and reliable security guarantees to protect DNNs against adversarial attacks. 

    \item To improve the resilience of DNNs against hardware-induced reliability threats, several low-cost fault-mitigation techniques have been proposed, e.g., range restriction-based fault mitigation~\cite{chen2020ranger}\cite{hoang2020ft}. These techniques have the potential of acting as a strong countermeasure against fault-injection attacks. However, their effectiveness as a countermeasure has not been studied so far. Alongside security, such studies can also help in further improving the reliability of DNN-based systems. 
\end{enumerate}
\section*{Acknowledgment}
This work is supported in parts by the Austrian Research Promotion Agency (FFG) and the Austrian Federal Ministry for Transport, Innovation, and Technology (BMVIT) under the ``ICT of the Future'' project, IoT4CPS: Trustworthy IoT for Cyber-Physical Systems.
%=====================================================================

\bibliographystyle{IEEEtran}
\bibliography{bib/bibliography}
%=====================================================================

\end{document}